# On the analytical solution in non-inertial frame of R2BP

**Sergey Ershkov***, Plekhanov Russian University of Economics,

Scopus number 60030998, e-mail: sergej-ershkov@yandex.ru,

**Dmytro Leshchenko**, Odessa State Academy of Civil Engineering and Architecture, Odessa, Ukraine, e-mail: leshchenko_d@ukr.net

## Abstract

In this analytical study, we have presented a new type of solving procedure with aim to obtain the coordinates of small mass *m*, which moves around primary $M_{Sun}$, referred to non-inertial frame of restricted two-body problem (R2BP) with modified potential function $U = -\frac{\mu}{R}$, $R = \sqrt{(x + \int V_1\, d\,t)^2 + (y + \int V_2\, d\,t)}$ (where $\{V_1(t),\, V_2(t)\}$ are the components of variable velocity $\vec{V}$ of central body $M_{Sun}$ motion) instead of classical potential function $U = -\frac{\mu}{R}$, $R = \sqrt{x^2 + y^2}$ for *Kepler's* formulation of R2BP. Meanwhile, system of equations of motion has been successfully explored with respect to the existence of analytical way for presentation of the solution in polar coordinates $X = x + \int V_1\, d\,t = r\cos\varphi$, $Y = y + \int V_2\, d\,t = r\sin\varphi$, $r = R$. We have obtained analytical formula for function $t = t(r)$ via appropriate elliptic integral. Having obtained the inversed dependence $r = r(t)$, we can obtain the time-dependence $\varphi = \varphi(t)$. Also, we have pointed out how to express components of solution (including initial conditions) from cartesian to polar coordinates as well.

## 1. Introduction, equations of motion.

In the restricted two-body problem (R2BP), the equations of motion describe the dynamics of a sufficiently small satellite $m$ under the action of gravitational force effected by one large celestial body $M_{Sun}$ ($m \ll M_{Sun}$). The small mass $m$ is supposed to move (as first approximation) inside the *restricted* region of space near the mass $M_{Sun}$ [1] without influencing of position of large celestial body $M_{Sun}$ even in anywhat negligible extent (but outside the Roche's limit [2] which is, as first approximation, not less than 7-10 $R_{Sun}$ where $R_{Sun}$ is the radius of the celestial body $M_{Sun}$). In case of *newtonian* type of gravitational forces, there is well-known analytical solution to the aforementioned problem (which has been associated earlier with Kepler's type of orbital motions both for the satellite and large celestial body around their common barycenter if we consider $m < M_{Sun}$ , instead of case $m \ll M_{Sun}$). It is also known from classical works that if large celestial body $M_{Sun}$ is in a fixed position in the problem under consideration or it is moving with constant velocity (i.e., its motion can be referred with respect to the inertial frame), the aforeformulated problem has the similar kepler-type solution. So, the main aim and motivation of this research concerns the investigation of more complicated case (than classical one) regarding existence of analytical solution in non-inertial case of R2BP where $\{V_1(t), V_2(t)\}$ are the components of observable *variable* velocity $\vec{V}$ of central body $M_{Sun}$ which is supposed to be moving all the time in one and the same direction but with *variable* velocity.

The problem of two bodies represents the core of celestial mechanical studies, as well as the starting point to strengthen our understanding of the n-body problem.

It is worth noting that there is a large number of previous and recent fundamental works concerning analytical generalization of the R2BP equations to the case of three or even many bodies, which should be mentioned accordingly [1-14]. We should especially emphasize the theory of orbits, which was developed in profound work [3] by V. Szebehely for the case of the circular restricted problem of three bodies (CR3BP) (primaries are rotating around their common centre of mass on *circular* orbits) as well as the case of the elliptic restricted problem of three bodies [4] (ER3BP, primaries are rotating around barycenter on *elliptic* orbits).

Let us consider here and below a non-rotating and *non-inertial* cartesian coordinate system with the origin $O$ located at the chosen initial moment $t_0$ in the center of mass of celestial body $M_{Sun}$ which moves strightly forward in one and the same direction (without rotation) with velocity $\vec{V} = \{V_1(t),\, V_2(t)\}$. Since transformation of velocity field from inertial coordinate system to the non-inertial frame of cartesian coordinate system $\vec{r}$ is expressed as follows [10, &39, p.166] (here below $\vec{\Omega}$ is pseudo-vector of the constant angular rotation)

$$\vec{v}_{inertial} = \vec{v}_{non-inertial} + \vec{V} + \vec{\Omega}\times\vec{r}\,, \quad \Rightarrow$$
$$\left(\frac{d\,\vec{v}_{inertial}}{d\,t}\right)_{inertial} = \left(\frac{d\,\vec{v}_{non-inertial}}{d\,t}\right)_{non-inertial} + \frac{d\vec{V}}{d\,t} + 2\,\vec{\Omega}\times\vec{v}_{non-inertial} + \vec{\Omega}\times(\vec{\Omega}\times\vec{r})\,, \tag{1}$$

so with help of (1) thus far the dynamical equations of motion for small mass $m$ with absence of rotation $\vec{\Omega} = \vec{0}$ can be written in well-known form as below [10, &39, p.166]

$$\begin{cases} \dfrac{d^2x}{d\,t^2} + \dfrac{d\,V_1}{d\,t} = -\dfrac{\partial U}{\partial\,x}\,, \\[2ex] \dfrac{d^2y}{d\,t^2} + \dfrac{d\,V_2}{d\,t} = -\dfrac{\partial U}{\partial\,y}\,, \end{cases} \tag{2}$$

where, $U$ is the potential function which should be determined as $U = -\dfrac{\mu}{R}$, $R = \sqrt{(x + \int V_1\,d\,t)^2 + (y + \int V_2\,d\,t)^2}$ (whereas $\{V_1(t),\, V_2(t)\}$ are the components of observable velocity of central body $M_{Sun}$ motion) instead of classical potential function $U = -\frac{\mu}{R}$, $R = \sqrt{x^2 + y^2}$ for *Kepler's* formulation of R2BP (here below and above, $\mu$ = const is the gravitational parameter in appropriate scale). Let us remark that partial derivatives in the right parts of Eqns. (2) should not be changed since expressions for $\{V_1(t), V_2(t)\}$ do not contain variables $\{x, y\}$ but depend only on time $t$.Initial conditions are as follows (dot indicates (d/d $t$) in (3)):

$$\left\{ \begin{array}{l} x(0) = 1,\ y(0) = 1, \\ \dot{x}(0) = \varepsilon = const\,(\varepsilon \sim 0), \\ \dot{y}(0) = \sqrt{1 - (\dot{x}(0))^2} \end{array} \right\} \tag{3}$$

## 2. Solving procedure for the system of Eqns. (2) with initial data (3).

Let us transform system (2) by the change of variables $X = x + \int V_1\, dt, \quad Y = y + \int V_2\, dt$

$$\left\{ \begin{array}{l} \dfrac{d^2 X}{dt^2} = -\dfrac{\mu X}{\left(X^2 + Y^2\right)^{\frac{3}{2}}}, \\ \\ \dfrac{d^2 Y}{dt^2} = -\dfrac{\mu Y}{\left(X^2 + Y^2\right)^{\frac{3}{2}}}. \end{array} \right. \tag{4}$$

Let us further transform system (4) by the change of variables $X = r{\cdot}\cos\varphi$, $Y = r{\cdot}\sin\varphi$ to the polar coordinates $\{r = r(t),\ \varphi = \varphi(t)\}$, $r = R = \sqrt{X^2 + Y^2}$, as below

$$\left(\frac{dX}{dt}\right) = r'\cos\varphi - r\sin\varphi\varphi', \qquad \left(\frac{dY}{dt}\right) = r'\sin\varphi + r\cos\varphi\varphi',$$

$$\frac{d^2 X}{dt^2} = r''\cos\varphi - 2r'\sin\varphi\varphi' - r\cos\varphi(\varphi')^2 - r\sin\varphi\varphi'',$$

$$\left(\frac{d^2 Y}{dt^2}\right) = r''\sin\varphi + 2r'\cos\varphi\varphi' - r\sin\varphi(\varphi')^2 + r\cos\varphi\varphi'', \qquad \Rightarrow$$

$$\begin{cases} r''\cos\varphi - 2r'\sin\varphi\varphi' - r\cos\varphi(\varphi')^2 - r\sin\varphi\varphi'' = -\dfrac{\mu\cos\varphi}{r^2}, \\ \\ r''\sin\varphi + 2r'\cos\varphi\varphi' - r\sin\varphi(\varphi')^2 + r\cos\varphi\varphi'' = -\dfrac{\mu\sin\varphi}{r^2}. \end{cases} \tag{5}$$

As first step, let us multiply first equation of the last system (5) onto cosφ, second onto sinφ, then sum the resulting equations one to each other:

$$r'' - r(\varphi')^2 = -\frac{\mu}{r^2} \quad \Rightarrow \quad (r\cdot(\varphi'))^2 = r\cdot r'' + \frac{\mu}{r} \tag{6}$$

The second step, let us multiply first equation of the last system onto sinφ, second onto cosφ, then subtract the resulting equations one from the other:

$$-2r'\varphi' - r\varphi'' = 0 \quad \Rightarrow \quad -\frac{2r'}{r} = \frac{\varphi''}{\varphi'} \quad \Rightarrow \quad -\frac{2\,d\,r}{r} = \frac{d\,(\varphi')}{\varphi'}$$

$$\Rightarrow \qquad \ln\left(\frac{r_0^2}{r^2}\right) = \ln\left(\frac{\varphi'}{\varphi'_0}\right) \quad \Rightarrow \quad \varphi' = \varphi'_0\cdot\left(\frac{r_0^2}{r^2}\right) \tag{7}$$

Taking into account (7), we could obtain from (6) as follows

$$r\cdot r'' - \frac{(\varphi'_0)^2\cdot r_0^4}{r^2} + \frac{\mu}{r} = 0 \quad \Rightarrow \quad \left\{ \frac{d\,r}{d\,t} = r' \equiv p(r) \Rightarrow r'' = \frac{d\,p}{d\,r}\cdot p \right\}$$

$$\Rightarrow \quad r\cdot\frac{d\,p}{d\,r}\cdot p - \frac{(\varphi'_0)^2\cdot r_0^4}{r^2} + \frac{\mu}{r} = 0 \quad \Rightarrow \quad \frac{1}{2}\frac{d\,(p^2)}{d\,r} = \frac{(\varphi'_0)^2\cdot r_0^4}{r^3} - \frac{\mu}{r^2}$$

$$(p^2 - (r'_0)^2) = 2(\varphi'_0)^2\cdot r_0^4\cdot(-\frac{1}{2})\cdot\left(\frac{1}{r^2} - \frac{1}{r_0^2}\right) + 2\mu\left(\frac{1}{r} - \frac{1}{r_0}\right) \quad \Rightarrow$$

then further after having obtained the quadrature in the left part of Eqn. (8) below (by appropriate approximation technique or e.g. by series of Taylor expansions)

$$p = \frac{d\,r}{d\,t} = \sqrt{(r'_0)^2 + 2(\varphi'_0)^2 \cdot r_0^4 \cdot (-\frac{1}{2}) \cdot \left(\frac{1}{r^2} - \frac{1}{r_0^2}\right) + 2\mu\left(\frac{1}{r} - \frac{1}{r_0}\right)} \quad \Rightarrow$$

$$\int \frac{d\,r}{\sqrt{(r'_0)^2 - (\varphi'_0)^2 \cdot r_0^4 \cdot \left(\frac{1}{r^2} - \frac{1}{r_0^2}\right) + 2\mu\left(\frac{1}{r} - \frac{1}{r_0}\right)}} = \pm \int d\,t \qquad (8)$$

$$\left\{ (r'_0)^2 - (\varphi'_0)^2 \cdot r_0^4 \cdot \left(\frac{1}{R^2} - \frac{1}{r_0^2}\right) + 2\mu\left(\frac{1}{R} - \frac{1}{r_0}\right) > 0 \right\}$$

we should find then the re-inverse dependence $r = r(t)$ (but since the power of polynomial under the sign of square root is much than 2, the left part of (8) presents the appropriate elliptic integral). Then afterwards we could obtain angle $\varphi$ by direct integration procedure, using (7).

## 3. <u>Discussion.</u>

As we can see from the derivation above, equations of motion (1) are proved to be very hard to solve analytically. Nevertheless, we have succeeded in obtaining analytical formulae for the components of the solution (6)-(8) in the polar coordinates $\{r(t), \varphi(t)\}$. Let us clarify that at transforming of equations (5) by virtue of special change of variables we have taken into account that independent variable (time $t$) is not included to the left and right parts of system (5). So, we have reduced this ordinary differential equation of 2-nd order (6) by the elegant change of variables $\left\{ \frac{dr}{dt} = r' \equiv p(r) \Rightarrow r'' = \frac{dp}{dr} \cdot p \right\}$ to the 1-st order differential equation. Then, having solved equation with regard to function $p(t)$, we should solve ODE in regard to $p = \frac{dr}{dt} = \sqrt{(r_0')^2 - (\varphi_0')^2 \cdot r_0^4 \cdot \left( \frac{1}{r^2} - \frac{1}{r_0^2} \right) + 2\mu \left( \frac{1}{r} - \frac{1}{r_0} \right)}$ to obtain the final result.

Ending discussion, let us note how to transform components of solution (6)-(8) from polar back → to cartesian coordinates (including initial conditions in general form). Quadrature (8) determines the dependence in general form $t = t(r)$, which contains the elliptic integral in the left part of (8) {under appropriate initial conditions; the upper limit of integral equals to $r$, low limit equals to $r_0$}, the right part of the quadrature (8) equals to $(t - t_0)$. We should re-inverse this expression into dependence $r = r(t)$, which can be obtained by numerical methods only (by appropriate approximation technique or e.g. by series of Taylor expansions).

Having obtained the dependence $r = r(t)$ from (8), we can then obtain from formula (7) the dependence (9) for $\varphi = \varphi(t)$:

$$\varphi = \varphi_0 + \varphi_0' \cdot \int_{t_0}^{t} \left( \frac{r_0^2}{r^2(t)} \right) dt \qquad (9)$$

Let us also recall that the change of variables $X = r \cdot \cos\varphi$, $Y = r \cdot \sin\varphi$ has been used at transformation of system (4). This fact means that the transformation of initial coordinates should be accordingly done as pointed out in (10-12) below:

$$r_0 = \sqrt{X_0^2 + Y_0^2}\,, \quad \varphi_0 = \arccos\left( \frac{X_0}{\sqrt{X_0^2 + Y_0^2}} \right),$$

$$\left\{ \left( \frac{dX}{dt} \right) = r' \cos\varphi - r \sin\varphi \varphi', \quad \left( \frac{dY}{dt} \right) = r' \sin\varphi + r \cos\varphi \varphi' \right\} \qquad (10)$$

$\Rightarrow$

$$1)\quad r' = \left(\frac{dX}{dt}\right)\cos\varphi + \left(\frac{dY}{dt}\right)\sin\varphi \quad \Rightarrow$$

$$r'_0 = \left(\frac{dX}{dt}\right)_0 \cdot \left(\frac{X_0}{\sqrt{X_0^2 + Y_0^2}}\right) + \left(\frac{dY}{dt}\right)_0 \sin\left(\arccos\left(\frac{X_0}{\sqrt{X_0^2 + Y_0^2}}\right)\right) \qquad (11)$$

$$2)\quad \left(\frac{dY}{dt}\right)\cos\varphi - \left(\frac{dX}{dt}\right)\sin\varphi = r\varphi' \quad \Rightarrow$$

$$\varphi'_0 = \frac{\left(\frac{dY}{dt}\right)_0 \left(\frac{X_0}{\sqrt{X_0^2 + Y_0^2}}\right) - \left(\frac{dX}{dt}\right)_0 \sin\left(\arccos\left(\frac{X_0}{\sqrt{X_0^2 + Y_0^2}}\right)\right)}{\sqrt{X_0^2 + Y_0^2}} \qquad (12)$$

## 4. **Conclusion.**

In this paper, we have presented a new type of the solving procedure to obtain the coordinates of the infinitesimal mass *m* which moves around the primary $M_{Sun}$ ($m \ll M_{Sun}$) for a special kind of restricted two-body problem, where $M_{Sun}$ moves with variable velocity $\vec{V} = \{V_1(t), V_2(t)\}$, with modified potential function $U = -\frac{\mu}{R}$, $R = \sqrt{(x + \int V_1\, dt)^2 + (y + \int V_2\, dt)}$ (where $\{V_1, V_2\}$ are the components of observable *variable* velocity of central body $M_{Sun}$ motion) instead of classical potential function $U = -\frac{\mu}{R}$, $R = \sqrt{x^2 + y^2}$ for *Kepler's* formulation of R2BP. Meanwhile, the system of equations of motion has been successfully explored with respect to the existence of analytical way for presentation of the solution in polar coordinates $X = x + \int V_1\, dt = r\cos\varphi$, $Y = y + \int V_2\, dt = r\sin\varphi$, $r = R$.

We have obtained analytical formula (8) for function $t = t(r)$. Having obtained the re-inverse dependence $r = r(t)$, we can obtain then the dependence $\varphi = \varphi(t)$ via formula (7). Also, we have pointed out how to express components of solution (including initial conditions) from cartesian to polar coordinates in general form (11)-(12).

The last but not least, we should especially note that such a kind of restricted two-body problem (presented in the current research) is found to be realistic for practical application in the real astophysical problems. Namely, when binary system (where large celestial body $M_{Sun}$ is the leading Primary Mover) is moving with observable but variable velocity $\vec{V} = \{V_1(t), V_2(t)\}$ ) towards another star system [11], such system will nevertheless keep *kepler-type* motion of secondary body $m << M_{Sun}$ around the primary body $M_{Sun}$.

## Acknowledgements

Sergey Ershkov is thankful to Prof. Nikolay Emelyanov for valuable comprehensive advices during fruitful discussions in the proccess of preparing of this manuscript.

## Conflict of interest

Authors declare that there is no conflict of interests regarding publication of article.

In this research, Dr. Sergey Ershkov is responsible for the general ansatz and the solving procedure, simple algebra manipulations, calculations, results of the article in Sections 1-3 and also is responsible for the search of approximated solutions.

Dr. Dmytro Leshchenko is responsible for theoretical investigations as well as for the deep survey in literature on the problem under consideration. Both authors agreed with the results and conclusions of each other in Sections 1-4.